\documentstyle[prl,aps,twocolumn,psfig]{revtex}
\draft
\begin{document}
\twocolumn[\hsize\textwidth\columnwidth\hsize
\csname @twocolumnfalse\endcsname
\title{Field induced transitions in a kagom\'e antiferromagnet}
\author{M. E. Zhitomirsky}
\address{European Synchrotron Radiation Facility, B.P.-220,
         Grenoble F-38043, France}
\date{August 1, 2001}
\maketitle
\begin{abstract}
The thermal order by disorder effect in magnetic field
is studied for a classical Heisenberg antiferromagnet on the kagom\'e
lattice. Using analytical arguments we predict a unique $H$--$T$
phase diagram for this strongly frustrated magnet: states with a coplanar
and a uniaxial triatic order parameters respectively at low and high
magnetic fields and an incompressible collinear spin-liquid state at
a one-third of the saturation field. We also present the Monte Carlo
data which confirm existence of these phases.
\end{abstract}
]

Geometrical frustration in lattice spin models is responsible for complete
suppression of conventional magnetic order and appearance of nonmagnetic
spin-liquids or states with exotic order parameters \cite{frust}. Applied
magnetic field can further enhance frustration. A well known example is
a weakly frustrated triangular lattice
antiferromagnet, which acquires an additional continuous degeneracy in
external field \cite{TAFM}. Investigation of high-field effects in
strongly frustrated magnets poses a new challenge for experimental
and theoretical studies \cite{GGG,ZHP}.

\begin{figure}[ht]
\centerline{\psfig{figure=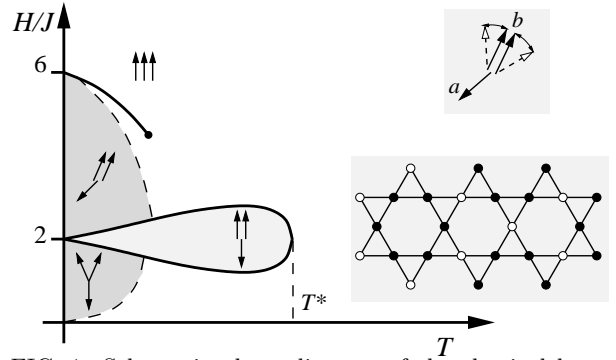,width=0.92\columnwidth}}
\caption{Schematic phase diagram of the classical kagom\'e antiferromagnet.
The top inset shows the clapping mode for quasicollinear states. The lower
inset shows a unit cell of the $\sqrt{3}\times\sqrt{3}$ quasicollinear
state: filled and empty circles denote the $b$- and the $a$-type spins,
respectively.}
\label{diagram}
\end{figure}

The Heisenberg antiferromagnet on the kagom\'e lattice (see Fig.~\ref{diagram})
is a strongly frustrated spin model, which is approximately realized in
a number of insulating layered magnets: SrCr$_8$Ga$_4$O$_{19}$ \cite{scgo}
and Ba$_2$Sn$_2$Ga$_3$ZnCr$_7$O$_{22}$ \cite{spinel} (both with $S=3/2$),
KFe$_3$(OH)$_6$(SO$_4$)$_2$ ($S=5/2$) \cite{jarosites}.
Gd$_3$Ga$_5$O$_{12}$ ($S=7/2$) is another frustrated magnet on a related
three-dimensional garnet lattice of corner-sharing triangles, which is often
called a hyper-kagom\'e lattice. This magnet has a weak exchange constant
$J\sim 1$~K and a peculiar unexplained phase diagram in magnetic field \cite{GGG}.
Motivated by the above materials with large values of spin we investigate in
this Letter the finite-temperature magnetization process of a classical
antiferromagnet on the kagom\'e lattice. We predict three distinctive field
regimes below the saturation field $H_{\rm sat}$, where exotic
spin phases are stabilized.

The Hamiltonian of a nearest-neighbor Heisenberg antiferromagnet
on the kagom\'e lattice with classical unit spins can be written
up to a constant term as:
\begin{equation}
\hat{\cal H}=\frac{1}{2}\sum_{\langle\bigtriangleup\rangle}^{N_\bigtriangleup}
\Bigl(J {\bf S}_\bigtriangleup^2 - {\bf H}\cdot {\bf S}_\bigtriangleup\Bigr) ,
\label{H}
\end{equation}
where the sum runs over all triangles, ${\bf S}_\bigtriangleup$ is the total spin
of a triangular plaquette and $N_\bigtriangleup=\frac{2}{3}N$ is the number of
plaquettes on an $N$-site lattice. The zero-field classical constraint
${\bf S}_\bigtriangleup\equiv 0$ fails to define a unique ground state.
The ground state coplanar configurations are constructed by fixing spins on a first
plaquette to a $120^\circ$-structure with left or right chirality for
the triad $\hat{\bf e}_a$, $\hat{\bf e}_b$, and $\hat{\bf e}_c$ and, then, tiling
this triad over the whole lattice in a way that three spins on every plaquette are
different \cite{kagome,weathervane,RB}. The number of all such states for
the kagom\'e lattice is known exactly: $U^N$ with $U=1.1347...$ \cite{baxter}.
Nonplanar ground states are constructed from planar configurations by identifying
so-called weathervane defects \cite{weathervane}.

The thermal order by disorder effect can appear 
because of a different entropy of short-wavelength fluctuations
above degenerate configurations \cite{obdo}. All coplanar states 
for a kagom\'e antiferromagnet show identical harmonic spectra. They have one 
flat zero-energy
branch with $N_4=\frac{1}{3}N$ modes, which corresponds to anharmonic quartic
excitations \cite{kagome,weathervane,RB}. A nonzero harmonic mode described
by a classical coordinate $y$ has energy increase
$\Delta E_2 \simeq Jy^2$ and contributes $\frac{1}{2}T\ln(J/T)$ 
to the thermodynamic potential, whereas a soft quartic mode with
$\Delta E_4 \simeq Jy^4$ 
makes a reduced contribution of $\frac{1}{4}T\ln(J/T)$. Coplanar configurations
have the largest number of soft modes and are, therefore, selected by
thermal fluctuations \cite{kagome}. Soft modes in coplanar states correspond
to alternate tilting of spins out of the ground state spin plane around
elementary hexagons. There are $\frac{1}{3}N$ hexagons on the kagom\'e lattice.
The counting of soft modes from such a geometrical point of view, thus, agrees with
the spin-wave analysis.

Harmonic fluctuations do not select between various planar 
configurations, though the highest statistical weight corresponds
to a so-called $\sqrt{3}\times\sqrt{3}$ structure \cite{RB}.
The low-temperature phase has nematic correlations of the chirality vectors
defined on every plaquette \cite{kagome}. Besides, the broken symmetries
include selection of the spin triad. The residual symmetry group 
is determined by the two elements: $C_6\hat{\cal T}$ and $U_2$, 
$C_6$ is a spin rotation by $\pi/3$ about normal to the spin plane,
$\hat{\cal T}$ is the time-reversal, and $U_2$ is a rotation by $\pi$ about
one of the three in-plane axes. The order parameter for low-temperature phases
of the kagom\'e antiferromagnet is a three-point correlation
function taken on one plaquette:
\begin{equation}
\langle S^\alpha({\bf r}_1)S^\beta({\bf r}_2)S^\gamma({\bf r}_3)\rangle
= S^{\alpha\beta\gamma} \ ,
\label{triatic}
\end{equation}
which has only a spin part described by a fully symmetric traceless tensor
$S^{\alpha\beta\gamma}$ \cite{weathervane,triatic}. For coplanar states at
$H=0$ this tensor is parameterized as
\begin{equation}
S^{\alpha\beta\gamma} = \sum_{a\leftrightarrow b\leftrightarrow c}
\hat{e}^\alpha_a\hat{e}^\beta_b\hat{e}^\gamma_c \ .
\label{circular}
\end{equation}
Note, that the three-spin (triatic) order parameter breaks explicitly
the time-reversal symmetry, which determines its nontrivial coupling to
an applied field.

In a finite magnetic field the classical energy Eq.~(\ref{H}) reaches
the minimum provided that
\begin{equation}
{\bf S}_\bigtriangleup = {\bf H}/(2J)
\label{constr}
\end{equation}
for every plaquette. Since
$|{\bf S}_\bigtriangleup|\leq 3$, above the saturation field $H_{\rm sat} = 6J$
all spins are aligned parallel to $\bf H$. We are interested in the field range
$0<H<H_{\rm sat}$, where the classical ground state possesses an infinite
degeneracy. The degeneracy is further enhanced by a field because the minimum
energy constraint does not require spins on one plaquette to lie in the same
plane. Shender and Holdsworth made the only attempt to understand
the finite field behavior of the kagom\'e antiferromagnet \cite{SH}.
They noticed that spin coplanarity responsible
for the increased number of soft modes is preserved if the spin plane
is parallel to an external field. Hence, thermal fluctuations create
magnetic anisotropy, which orients at low temperatures the spin plane.

There is an additional effect missed in the above consideration
and related to the triatic nature of the spin order parameter.
Once selection of the spin plane takes place, an extra {\it macroscopic}
degree of freedom appears: orientation of the spin triad inside the plane
relative to $\bf H$. At $T=0$ all orientations have the same
classical energy, though the spin triad distorts differently, when
it forms different angles $\varphi$ with the magnetic field.
This distortion modifies harmonic fluctuations
inside the spin plane. In such cases with a one-parameter
continuous degeneracy thermal fluctuations always
produce the order by disorder effect and select a homogeneous
state for one particular value of the parameter \cite{obdo}.
The orientational field effect for the triatic order parameter
is described by an invariant
$E_{\rm an}\simeq H^\alpha H^\beta H^\gamma S^{\alpha\beta\gamma}
\propto H^3\cos 3\varphi$, which is present in the Landau functional 
because of a broken time-reversal symmetry in the triatic phase.
The minimum energy is reached for $\varphi=0$ or $\pi$ depending
on a sign of the prefactor, i.e., one of the triad vectors has
to be either parallel or antiparallel to the applied field.

To find the equilibrium orientation of the spin triad in external field,
we performed spin-wave calculations. The harmonic spectrum of a coplanar
state at $H\neq 0$ depends not only on the triad orientation but also
on a way the triad is tiled over the lattice. We used, therefore, the
$\sqrt{3}\times\sqrt{3}$ and  the $q=0$ structures \cite{RB} as
two example cases. For both structures $\varphi=\pi$ is selected
by fluctuations with $E_{\rm an} = 5.0\,T(H/H_{\rm sat})^3N\cos 3\varphi$
and $1.3\,T(H/H_{\rm sat})^3N\cos 3\varphi$, respectively. Thus,
thermal fluctuations in a weak magnetic field
create anisotropy for the spin plane as well as inside the plane.
The kagom\'e antiferromagnet at low fields is described by the coplanar
triatic order parameter (\ref{circular}) with one of the spin triad
vectors being antiparallel to the field direction, Fig.~\ref{diagram}.

Generally, classical ground states at an arbitrary field
$0<H<H_{\rm sat}$ are all noncollinear. Collinear configurations appear
among the ground state manifold only at special rational values
of the applied field. The up-up-down ($uud$) structure has
$|{\bf S}_\bigtriangleup|=1$ and, hence, is stable only at
$H_c=2J=\frac{1}{3}H_{\rm sat}$.
The problem of calculating the total number of the $uud$
states for the kagom\'e lattice at $H=H_c$ can be mapped onto the problem
of dimer coverings of a dual hexagonal lattice,
which is solved exactly \cite{baxter}. The number
of $uud$ states scales as $V^N$, with $V=1.1137$...
A special role of such collinear spin configurations in magnetic field for
various frustrated models has been recently emphasized in \cite{ZHP}.
Standard harmonic analysis for the kagom\'e antiferromagnet at
$H=\frac{1}{3}H_{\rm sat}$ yields three two-fold degenerate
excitation branches for an arbitrary $uud$ state:
\begin{equation}
\omega^1_{\bf k} =0\ ,\ \
\omega^{2,3}_{\bf k} = 3\pm \sqrt{3(1+2\gamma_{\bf k})} \ ,
\label{freq}
\end{equation}
with $\gamma_{\bf k} = \frac{1}{3}(\cos k_x +
2\cos\frac{1}{2}k_x\cos\frac{\sqrt{3}}{2}k_y)$. Therefore, the collinear states
have $N_4=\frac{2}{3}N$ soft quartic modes, twice more than any
coplanar state. This fact has a simple geometrical origin. Local soft modes
correspond to alternate tilting of spins
around elementary hexagons. For collinear states such distortions
have two polarizations in two directions perpendicular to the field,
while coplanar states have soft modes only in the polarization
transverse to the spin plane.

Thermal fluctuations reduce the free energy of the $uud$ states
compared to other noncollinear classical ground states
at $H=\frac{1}{3}H_{\rm sat}$. Further lifting of degeneracy
within the discrete subset of the $uud$ states does not occur due to
the same mechanism as for coplanar states at $H=0$ \cite{kagome,weathervane,SH}.
Every collinear state has special lines of weathervane defects, which contain
alternating sequences of $u$-$d$-$u$-$d$ spins. Flipping all spins along such a line
by $180^\circ$ produces another collinear state at the free energy cost $\Delta F\simeq T$.
A macroscopic temperature independent number of such defects is generated
at low $T$'s leaving all $uud$ states accessible for the magnet and creating
a collinear spin liquid. Enhanced number of soft modes is also reflected
in the specific heat $C$: out of $2N$ degrees of freedom for
$N$ classical spins, every quadratic mode contributes $\frac{1}{2}k_B$
to the specific heat, while a quartic mode does
only $\frac{1}{4}k_B$ \cite{kagome}. Hence, $C = \bigl[1-N_4/(4N)\bigr] k_B$
per one spin and we expect $C=\frac{5}{6}k_B$ for the collinear spin liquid
compared to $C=\frac{11}{12}k_B$ for coplanar states. Specific heat measurement
in numerical simulations provides a useful way
to distinguish different spin states.

The free energy for the collinear spin liquid is lower by 
$\sim NT\ln(J/T)$ compared to any noncollinear state. But a small
deviation of field away from $H=\frac{1}{3}H_{\rm sat}$ produces
only a little change in the classical energy. The collinear spin liquid is,
therefore, stabilized by anharmonic interactions to a finite field range in a way
similar to stabilization of the incompressible quantum fluid states in
the Fractional Quantum Hall Effect.
The collinear spin liquid phase yields a {\it weak} magnetization
plateau at $1/3$ of the saturated
magnetization: spins in the $uud$ states cannot tilt towards the field
and a weak variation of the magnetization for the collinear spin liquid
is produced by thermal excitations alone. The above arguments also
suggest the first-order transition from the collinear spin liquid state
to the low- and the high-field phases at low temperatures.

In the field range $\frac{1}{3}H_{\rm sat}<H<H_{\rm sat}$ the ground state
configurations are again all noncollinear. There appear, however, new
quasicollinear states with two coinciding spins of the basis triad
$\hat{\bf e}_b=\hat{\bf e}_c$, which have zero chirality on every plaquette
(Fig.~\ref{diagram}). These configurations belong to the subset of
coplanar states. Hence, a quasicollinear state has all the soft excitations
of coplanar states: $\frac{1}{3}N$ quartic modes in the polarization
transverse to the spin plane. In addition, they acquire extra in-plane soft modes.
For a single plaquette, such a clapping-type mode corresponds to simultaneous tilting
of the two parallel spins in opposite directions inside the spin plane
(Fig.~\ref{diagram}). To preserve a weak anharmonic energy change the clapping
deformation has to be extended on the lattice along a loop which contains only
the $b$-type spins. The shortest such loop is a perimeter of a hexagon, while number of
appropriate hexagons is maximized 
if a quasicollinear state is tiled in the $\sqrt{3}\times\sqrt{3}$ structure
(Fig.~\ref{diagram}). Thus, the $\sqrt{3}\times\sqrt{3}$ quasicollinear state has
the largest number of soft modes $N_4 = \frac{4}{9}N$ at
$\frac{1}{3}H_{\rm sat}<H<H_{\rm sat}$. 

Nevertheless, the true breaking of the translational
symmetry does not take place. The disordering mechanism is again related
to the presence of loops of alternating spin orientations or weathervane defects
\cite{weathervane}. Loop flipping $a$-$b$-$a$-$b\rightarrow b$-$a$-$b$-$a$
at zero energy cost contributes $\sim T\ln(J/T)$ to the thermodynamic potential
because of a destruction of a few soft modes. Such losses are outweighed
at small concentrations of defects by the entropy gain. At finite
temperatures there will be a macroscopic number of defects, which
destroy the long-range $\sqrt{3}\times\sqrt{3}$ translational symmetry and
create instead a phase with a uniaxial triatic order parameter:
\begin{equation}
S^{\alpha\beta\gamma} = n^\alpha n^\beta n^\gamma - \case{1}{4}\bigl(
n^\alpha\delta_{\beta\gamma} + n^\beta\delta_{\gamma\alpha} +
n^\gamma\delta_{\alpha\beta} \bigr) \ ,
\label{uniaxial}
\end{equation}
where indices run over $x,y$ and ${\bf n}=\hat{\bf e}_a-
(\hat{\bf e}_a\cdot\hat{\bf z})$. Higher free energy costs make
the number of defects in the uniaxial triatic state to be
temperature dependent: $N_{\rm def} \sim TN$ with diverging
magnetic correlation length $\xi \sim 1/T$.
Lacking a conventional magnetic order the high-field triatic phase
asymptotically approaches the ordered $\sqrt{3}\times\sqrt{3}$ quasicollinear
structure. This behavior is reflected in the temperature dependence of
the specific heat which should tend to $C=\frac{8}{9}k_B$ at low $T$.

\begin{figure}[ht]
\centerline{\psfig{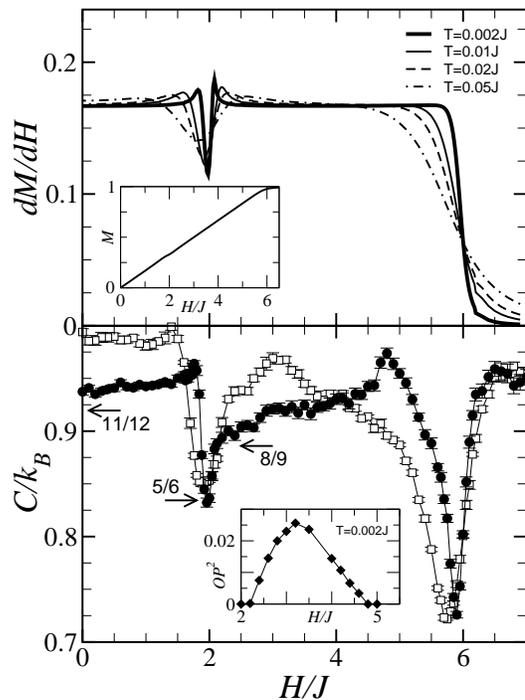}}
\caption{MC results for the classical kagom\'e antiferromagnet.
Top panel: susceptibility vs.\ magnetic field for several temperatures.
The inset shows the magnetization curve at $T=0.01J$. Bottom panel: specific
heat vs.\ magnetic field for $T=0.002J$ (full circles) and $T=0.01J$
(open squares). The inset shows the finite-size extrapolation for the
squared triatic order parameter (\ref{uniaxial}).}
\label{MCdata}
\end{figure}

At the saturation field $H_{\rm sat}=6J$ the ground state becomes nondegenerate
and corresponds to parallel alignment of spins. The harmonic excitations in
this collinear $uuu$ phase are given by the same expressions (\ref{freq}) as
for the $uud$ states. There are $N_4=\frac{2}{3}N$ soft modes at this magnetic field
and the specific heat again reaches the value $C=\frac{5}{6}k_B$. Soft modes also
produce a universal magnetic behavior of the frustrated spin system at
the saturation. For $H>H_{\rm sat}$ all three branches (\ref{freq}) acquire
an additional field dependent shift $\delta\omega = (H-H_{\rm sat})$. Taking into
account only the most singular contribution of the first flat branch, we derive that
at $H\ge H_{\rm sat}$ the classical partition function scales as
\begin{equation}
Z(H,T) = T^{1/4} f(u)\ , \ \ \ u=(H-H_{\rm sat})/\sqrt{T}\ .
\label{Z}
\end{equation}
The susceptibility $\chi=dM/dH$ determined from (\ref{Z}) is
$\chi = [f''(u)/f(u)]-[f'(u)/f(u)]^2$. At $u=0$ ($H=H_{\rm sat}$)
the susceptibility becomes temperature independent and all the curves $\chi(H,T)$
cross at one point, see Fig.~\ref{MCdata}.

So far we considered only the short wave-length degrees of freedom
and completely disregarded long wave-length excitations, which determine
the nature of phase transitions in two dimensions. The non-Abelian topological
defects discussed previously for the triatic order parameter
\cite{weathervane} are suppressed by the field.  The two triatic phases break
only $U(1)$ symmetry for rotations about the field direction.
Berezinskii-Kosterlitz-Thouless type transitions separate them from
the high temperature paramagnetic phase, see Fig.~\ref{diagram}.
In contrast, the first-order transition to the collinear spin-liquid state survives
and corresponds to a gas-liquid type
transition on the boundary with the paramagnetic phase.
The collinear $uuu$ phase also has an enhanced entropy contribution
and is stabilized below $H_{\rm sat}$.
The corresponding line of first-order transitions ends at a critical point
as sketched in Fig.~\ref{diagram}.

We have also performed Monte Carlo (MC) simulations for the model (\ref{H}).
The standard Metropolis algorithm was used with up to $10^7$ MC steps
per every point. Simulations were done for periodic $3L^2$-site lattices
with $L=6$, 12, 18; the data presented in Fig.~\ref{MCdata} are for a
972-spin cluster. The magnetization curve at $T=0$ 
is a straight line with a slope $\chi=dM/dH=\frac{1}{6}$.
At low temperatures $\chi(H)$ develops a dip near $H=\frac{1}{3}H_{\rm sat}$
corresponding to a weak plateau at $\frac{1}{3}M_{\rm sat}$. Two peaks,
which surround the dip, indicate first-order transitions to the plateau phase.
The peaks become rounded and, then, completely disappear at higher temperatures.
We estimated $T^*\simeq 0.025J$ as the highest temperature, where the collinear
spin-liquid still exists (Fig.~\ref{diagram}). Above this temperature the dip
corresponds to a smooth crossover. The field dependence of the specific heat also
follows the above predictions. At $T=0.002J$, the specific heat starts near
$\frac{11}{12}k_B$, corresponding to the coplanar triatic state, at low fields. As
the field approaches $\frac{1}{3}H_{\rm sat}$, $C(H)$ goes down to $\frac{5}{6}k_B$,
corresponding to the collinear spin liquid, and, then, recovers back to
$\frac{8}{9}k_B$, corresponding to the uniaxial triatic state. The specific heat
drops again near the saturation with $C|_{H_{\rm sat}}=\frac{5}{6}k_B$, while the
minimum value is reached at somewhat lower fields. Such a behavior signals
stabilization of the $uuu$ phase beyond its classical boundary: 
bare negative modes below $H_{\rm sat}$ are renormalized
into soft modes, which further decrease $C(H)$. The specific heat also exhibits
a peak at $H\approx 4.8J$ ($T=0.002J$) and $H\approx 3J$ ($T=0.01J$) with a
significant finite-size dependence. We attribute this peak to a phase transition
from the collinear $uuu$ state into a noncollinear uniaxial triatic state
(\ref{uniaxial}). To check this we have calculated field dependence of the squared
triatic order parameter (\ref{triatic}) at $T=0.002J$ for three cluster sizes and
extrapolated it to the thermodynamic limit. Results are shown in the inset in
Fig.~\ref{MCdata}. The uniaxial triatic phase disappears at about the same field
as the position of the peak in the specific heat. The present data do not 
resolve clearly the order of the phase transition, which must be
determined in a detailed numerical study.

We have shown that the geometrical approach based on soft mode counting
is a powerful tool for investigation of the field behavior of a
classical kagom\'e lattice antiferromagnet. 
Our preliminary analysis shows that a Heisenberg antiferromagnet
on the related garnet lattice has the same type
of the phase diagram, Fig.~\ref{diagram}.
In particular, the field range of an asymptotically ordered quasicollinear state 
coincides roughly with a region for a field-induced ordering
observed in Gd$_3$Ga$_5$O$_{12}$ \cite{GGG}.

I am grateful to J.\,T.\,Chalker, E.\,I.\,Kats, V.\,I.\,Mar\-chenko,
O.\,A.\,Petrenko, and J.\,D.\,M.\,Champion for useful discussions and
to A.\,A.\,Snigireva for helpful remarks.


\begin{thebibliography}{99}

\bibitem{frust}
 {\it Magnetic system with competing interactions\/}, edited by H.T. Diep
 (World Scientific, Singapore, 1994); P. Schiffer and A.P. Ramirez,
 Comments Cond. Mat. Phys. {\bf 18}, 21 (1996).

\bibitem{TAFM}
 D.H. Lee {\it et al\/}., Phys. Rev. Lett. {\bf 52}, 433 (1984);
 H. Kawamura, J. Phys. Soc. Jpn. {\bf 53}, 2452 (1984);
 S.E. Korshunov, J. Phys. C. {\bf 19}, 5927 (1986).

\bibitem{GGG}
 S. Hov {\it et al\/}., J. Magn. Magn. Mat. {\bf 15-18}, 455 (1980);
 P. Schiffer {\it et al}., Phys. Rev. Lett. {\bf 73}, 2500 (1994).
 O.A. Petrenko {\it et al}., {\it ibid\/}.  {\bf 80}, 4570 (1998).

\bibitem{ZHP}
 M.E. Zhitomirsky, A. Honecker, and O.A. Petrenko,
 Phys. Rev. Lett. {\bf 85}, 3269 (2000).

\bibitem{scgo}
 X. Obrados {\it et al\/}., Solid State Commun. {\bf 65}, 189 (1988);
 A.P. Ramirez {\it et al\/}., Phys. Rev. Lett. {\bf 64}, 2070 (1990).

\bibitem{spinel}
 I.S. Hagemann {\it et al\/}., Phys. Rev. Lett. {\bf 86}, 894 (2001).

\bibitem{jarosites}
 A.S. Wills {\it et al\/}., Phys. Rev. B {\bf 61}, 6156 (2000).

\bibitem{kagome}
 J.T. Chalker, P.C.W. Holdsworth, and E.F. Shender,
 Phys. Rev. Lett. {\bf 68}, 855 (1992).

\bibitem{weathervane}
 I. Ritchey, P. Chandra, and P. Coleman,
 Phys. Rev. B {\bf 47}, 15342 (1993).

\bibitem{RB}
 J.N. Reimers and A.J. Berlinsky, Phys. Rev. B {\bf 48}, 9539 (1993).

\bibitem{baxter}
 R.J. Baxter, J. Math. Phys. {\bf 11}, 784 (1970).

\bibitem{obdo}
 J. Villain {\it et al\/}., J. Phys. (Paris) {\bf 41}, 1263 (1980);
 E.F. Shender, Sov. Phys. JETP {\bf 56}, 178 (1982).

\bibitem{triatic}
 V.I. Marchenko, JETP Lett. {\bf 48}, 427 (1988).

\bibitem{SH}
 E.F. Shender and  P.C.W. Holdsworth,
 in {\it Fluctuations and Order\/}, ed.\ by M. Millonas
 (Springer, Berlin, 1995).

\end{thebibliography}
\end{document}